\documentclass[11pt]{article}

\usepackage{indentfirst}
\usepackage[utf8]{inputenc}
\usepackage{bm}
\usepackage{amsmath, amssymb, amsthm}
\usepackage{graphicx}
\usepackage{natbib}
\usepackage{geometry}
\usepackage{setspace}
\usepackage{hyperref}
\usepackage{booktabs}
\usepackage{algorithm}
\usepackage{algorithmic}
\usepackage{caption}
\usepackage{float}     
\usepackage{geometry}  
\usepackage{natbib}
\hypersetup{colorlinks=true, linkcolor=blue, citecolor=blue, urlcolor=blue}

\title{Empirical Validation of Functional Multidimensional Scaling via Numerical Simulation and Real-World Application}
\author{Liting Li \\
Department of Mathematical Sciences \\
University of Wisconsin-Milwaukee
}
\date{May 2025}

\begin{document}

\maketitle
\newpage
\begin{abstract}
This article presents an empirical validation of the functional multidimensional scaling model, a novel approach that improves the smoothness of time-varying dissimilarities in a low-dimensional space, embedding a modified Adam stochastic gradient descent method. We conduct a numerical simulation study to evaluate the feasibility of the functional multidimensional scaling model under various controlled scenarios and to assess the goodness of the approximation of the estimators with a curvilinear search method, demonstrating its robustness and scalability in dynamic structures. To further explore its effectiveness in practice, we implement the functional multidimensional scaling model in a real-world case with stock market data, revealing strong clustering capabilities in visualization. The experiments in this article indicate that the functional multidimensional scaling model performs robustly on synthetic benchmarks and provides meaningful insights of high-dimensional and time-varying data in the real world, reinforcing its value in practical applications.

\textit{Keywords:} Functional Multidimensional Scaling, Time-Varying Dissimilarity, Multivariate Normal Distribution, Adam Stochastic Gradient Descent, Barzilai-Borwein Step Size
\end{abstract}

\newpage

\section{Introduction}
\noindent Multidimensional time-varying data is prevalent in areas such as climate science, marketing, and finance. Classical multidimensional scaling (CMDS) in \cite{ramsay2005functional} was developed as one of the most widely used multidimensional scaling (MDS) techniques for visualizing multivariate data by preserving pairwise dissimilarities among objects in a low-dimensional representation. CMDS offers a mathematical simplicity and an interpretable solution for static data, by transforming a dissimilarity matrix into a set of coordinates via eigendecomposition. While CMDS has conceptual simplicity and effectiveness in constructing a low-dimensional space to illustrate relationships between objects, it assumes fixed dissimilarity among objects and lacks adaptability to changing temporal patterns, which limits its utility in dynamic environment, such as financial market, marketing strategies, or climate systems. To leverage the extension in dynamic MDS, several studies have explored temporal multidimensional scaling in various domains, for example, \cite{machado2011analysis} conducted MDS analysis in 15 stock markets over the world with their time-varying correlations. Moreover, in network security area, \cite{Jackle2016TMDS} created one-dimensional MDS plots for multivariate time series data. Although all of the research makes a significant breakthrough in time-varying MDS, little attention has been paid to the smoothness of MDS representations if dissimilarities of objects are considered to change over time.

In our previous work (\cite{li2025functional}), we proposed functional multidimensional scaling (FMDS), a novel MDS model that incorporates the smoothness of functional data analysis and the interpretability of multidimensional scaling techniques. As a variant of MDS, FMDS is especially suited for time-varying dissimilarities, enabling continuous monitoring of structural changes and offering smooth clustering capabilities in dynamic systems by applying B-spline basis functions to reduce random noise in consecutive data points. To mitigate the computational burden arising from multidimensional time-varying data, FMDS implements an embedding optimization with a modified stochastic gradient descent method. This efficient optimization strategy improves the scalability and stability of MDS while preserving dissimilarities in dynamic data structure.

The key contributions of this article are composed of four sections. Firstly, we briefly overview the theoretical background in the FMDS model (Section 2). In order to measure the consistency performance of FMDS as time changes, we conduct a numerical simulation experiment with synthetic data (Section 3). Subsequently, we demonstrate the empirical validation of FMDS by applying it to a real-world case with stock price data of S\&P 500 Index, showcasing the efficiency of the modified stochastic gradient descent method and validating the smoothness of the FMDS model with time-varying dissimilarities among the stocks (Section 4). The FMDS model also exhibits 2-D plots for the stocks that can be used to analyze the relationships of them over the year 2018. In the final section of the article (Section 5), we discuss the challenges and future directions of the applications of FMDS.
\newpage
\section{Background on Functional Multidimensional Scaling}
\noindent Until now, no large-scale studies have been performed to investigate multidimensional scaling (MDS) solutions with time-varying dissimilarities. The work made by \cite{ambrosi1987dynamischer} is one of the few literature about time-varying MDS. They developed a dynamic approach to produce an MDS reconstruction of $n$ objects, given their dissimilarities measured over a consecutive time period. Their study is based on the loss function for nonmetric MDS that considers qualitative dissimilarities. Given $d_{ij}^t$ as the dissimilarity between objects $i$ and $j$ at the time point $t$, where $t = 1,...,T$, the dissimilarity $\hat{d}_{ij}^t$ among the optimal dynamic MDS solutions $\mathbf{\hat x}_i^t = [x_{i1}^t, ... ,x_{ip}^t]^\top \in \mathbb{R}^p$ and $\mathbf{\hat x}_j^t = [x_{j1}^t, ... ,x_{jp}^t]^\top \in \mathbb{R}^p$ is obtained by minimizing the overall loss function with a penalty function for the entire period of time. However, this method causes drastic changes during the time period for each object and lacks smoothness in the MDS representations.

To overcome this challenge, we built upon the functional data analysis and proposed the functional multidimensional scaling (FMDS) model (\cite{li2025functional}) that contributes smoothness to time-varying MDS solutions. Assume that there are $n$ objects $\mathbf{y}_1(t), \mathbf{y}_2(t),...,\mathbf{y}_n(t) \in \mathbb{R}^r$ at the time point $t$, and the dissimilarity between the objects $i$ and $j$ is denoted by $d_{ij}(t)$. To be clarified, $d_{ij}(t)$ is not a function of $t$ but the discrete dissimilarity at the time point $t$. In order to obtain smoothly time-varying representations, we proposed FMDS with B-spline basis functions that produces the functional MDS solutions $\mathbf{x}_1(t), \mathbf{x}_2(t),..., \mathbf{x}_n(t)$ such that 
\begin{center}
$\|\mathbf{x}_i(t) - \mathbf{x}_j(t)\| \approx d_{ij}(t)$, for $i,j=1,...,n$,
\end{center} 

Considering that the smoothness characteristic of B-spline basis functions contribute to functional data introduced by \cite{ramsay2005functional}, we let $\mathbf{x}_i(t) = \mathbf{C}_i \boldsymbol{\beta}(t)$, where $\boldsymbol{\beta}(t)$ is the vector of $q$ B-spline basis functions and each $\mathbf{C}_i$ is a $(p \times q)$ coefficient matrix, for $i=1,...,n$. The estimated $\mathbf{\hat C}_1,...,\mathbf{\hat C}_n$ are motivated by the target function
\begin{align*}
F(\mathbf{C}_1,...,\mathbf{C}_n) &= \sum_{i<j}^{n} \sum_{k=1}^{m} [d_{ij}^2(t_k) - \|\mathbf{x}_i(t_k) - \mathbf{x}_j(t_k)\|^2]^2 \\
 &= \sum_{i<j}^{n} \sum_{k=1}^{m} [d_{ij}^2(t_k) - \|\mathbf{C}_i \boldsymbol{\beta}(t_k) - \mathbf{C}_j \boldsymbol{\beta}(t_k)\|^2]^2 \\
 &= \sum_{i<j}^{n} \sum_{k=1}^{m} [d_{ij}^2(t_k) - (\mathbf{C}_i \boldsymbol{\beta}(t_k) - \mathbf{C}_j \boldsymbol{\beta}(t_k))^\top (\mathbf{C}_i \boldsymbol{\beta}(t_k) - \mathbf{C}_j \boldsymbol{\beta}(t_k))]^2, \tag{2.1}
\end{align*}
where $m$ is the length of time period with a continuous interval $\tau$. Consequently, we have the partial derivatives of $\mathbf{C}_h$ as follows, \\
\begin{align*}
\frac{\partial F}{\partial{\mathbf{C}_h}} &= \sum_{\substack{j=1 \\ j \neq h}}^{n} \sum_{k=1}^{m} \frac{\partial}{\partial \mathbf{C}_h}[d_{hj}^2(t_k) - (\mathbf{C}_h \boldsymbol{\beta}(t_k) - \mathbf{C}_j \boldsymbol{\beta}(t_k))^\top (\mathbf{C}_h \boldsymbol{\beta}(t_k) - \mathbf{C}_j \boldsymbol{\beta}(t_k))]^2 \\
  &= -4\sum_{\substack{j=1 \\ j \neq h}}^{n} \sum_{k=1}^{m} [d_{hj}^2(t_k) - (\mathbf{C}_h \boldsymbol{\beta}(t_k) - \mathbf{C}_j \boldsymbol{\beta}(t_k))^\top (\mathbf{C}_h \boldsymbol{\beta}(t_k) - \mathbf{C}_j \boldsymbol{\beta}(t_k))](\mathbf{C}_h \boldsymbol{\beta}(t_k) - \mathbf{C}_j \boldsymbol{\beta}(t_k)) \boldsymbol{\beta}(t_k)^\top
\end{align*}
In order to estimate the coefficient matrices $\mathbf{C}_1,...,\mathbf{C}_n$, we minimize the target function by letting the partial derivatives be zero, that is, 
\begin{align*}
\frac{\partial F}{\partial\mathbf{C}_h} = \mathbf{0}. 
\end{align*}
Hence, the minimization can be obtained by
\begin{center}
\[
\sum_{\substack{j=1 \\ j \neq h}}^{n} \sum_{k=1}^{m} [d_{hj}^2(t_k) - (\mathbf{C}_h \boldsymbol{\beta}(t_k) - \mathbf{C}_j \boldsymbol{\beta}(t_k))^\top (\mathbf{C}_h \boldsymbol{\beta}(t_k) - \mathbf{C}_j \boldsymbol{\beta}(t_k))](\mathbf{C}_h \boldsymbol{\beta}(t_k) - \mathbf{C}_j \boldsymbol{\beta}(t_k)) \boldsymbol{\beta}(t_k)^\top = 0
\]
\end{center}
However, each $\mathbf{C}_h$ cannot be solved in a closed form from the above equation. To solve out each $\mathbf{C}_h$ from this high-dimensional problem that requires substantial computation time, our previous work developed a modified Adam stochastic gradient descent (SGD) method, motivated by \cite{kingma2015adam} which only considers first-order gradients of individual sub-function with little memory requirement. 

In the target function (2.1), the individual sub-functions are given by 
\begin{align*}
f(\mathbf{C}_h, \mathbf{C}_j) &= \sum_{k=1}^{m}[d_{hj}^2(t_k) - \|\mathbf{x}_h(t_k) - \mathbf{x}_j(t_k)\|^2]^2 \\
  &= \sum_{k=1}^{m} [d_{hj}^2(t_k) - (\mathbf{C}_h \boldsymbol{\beta}(t_k) - \mathbf{C}_j \boldsymbol{\beta}(t_k))^\top (\mathbf{C}_h \boldsymbol{\beta}(t_k) - \mathbf{C}_j \boldsymbol{\beta}(t_k))]^2,
\end{align*}
which results in the first-order stochastic gradient functions with respect to $\mathbf{C}_h$ and $\mathbf{C}_j$ ($h \neq j$),
\begin{align*}
\frac{\partial f}{\partial \mathbf{C}_h} &= -4\sum_{k=1}^{m} [d_{hj}^2(t_k) - (\mathbf{C}_h \boldsymbol{\beta}(t_k) - \mathbf{C}_j \boldsymbol{\beta}(t_k))^\top (\mathbf{C}_h \boldsymbol{\beta}(t_k) - \mathbf{C}_j \boldsymbol{\beta}(t_k))][\mathbf{C}_h \boldsymbol{\beta}(t_k) - \mathbf{C}_j \boldsymbol{\beta}(t_k)] \boldsymbol{\beta}(t_k)^\top, \\
\frac{\partial f}{\partial \mathbf{C}_j} &= 4\sum_{k=1}^{m} [d_{hj}^2(t_k) - (\mathbf{C}_h \boldsymbol{\beta}(t_k) - \mathbf{C}_j \boldsymbol{\beta}(t_k))^\top (\mathbf{C}_h \boldsymbol{\beta}(t_k) - \mathbf{C}_j \boldsymbol{\beta}(t_k))][\mathbf{C}_h \boldsymbol{\beta}(t_k) - \mathbf{C}_j \boldsymbol{\beta}(t_k)] \boldsymbol{\beta}(t_k)^\top.
\end{align*}
In our modified Adam SGD method, we initialize $\mathbf{C}_1,...,\mathbf{C}_n$ by using CMDS with the given dissimilarities $d_{ij}$'s, update each pair of $\mathbf{C}_h$ and $\mathbf{C}_j$ with random sample $h$ selected from $\{1,...,n-1\}$ and $j=h+1,h+2,...,n$, respectively in each iteration $i$. The iteration stops when $\mathbf{C}_1,...,\mathbf{C}_n$ tend to converge. The algorithm is described in Algorithm 1.

\begin{algorithm}[H]
\caption{Modified Adam Stochastic Gradient Descent for FMDS}
\label{alg:modified_adam}
\begin{algorithmic}[1]
\REQUIRE Learning rate $\alpha = 0.001$, exponential decay rates $\gamma_1 = 0.9$, $\gamma_2 = 0.999$, \\
$\mathbf{e}_h$ and $\mathbf{e}_j$ are both $p \times q$ constant matrices with all elements equal to $10^{-8}$, \\
initial parameters $\mathbf{C}_h$ and $\mathbf{C}_j$ by using CMDS with the given dissimilarities $d_{ij}$'s, \\
$h$ is a random sample selected from $\{1,...,n-1\}$ and $j=h+1,h+2,...,n$.
\ENSURE Optimized each pair of parameters $\mathbf{C}_h$ and $\mathbf{C}_j$
\STATE Initialize $\mathbf{m}_h^{(0)} = \mathbf{0}$, $\mathbf{v}_h^{(0)} = \mathbf{0}$, $\mathbf{m}_j^{(0)} = \mathbf{0}$, $\mathbf{v}_j^{(0)} = \mathbf{0}$ are $p \times q$ matrices, iteration step $i = 0$
\WHILE{$\|\mathbf{C}_h^{(i+1)} - \mathbf{C}_h^{(i)}\| \geq \epsilon$ and $\|\mathbf{C}_j^{(i+1)} - \mathbf{C}_j^{(i)}\| \geq \epsilon$}
    \STATE Compute gradient $\frac{\partial f}{\partial \mathbf{C}_h^{(i)}}$ and $\frac{\partial f}{\partial \mathbf{C}_j^{(i)}}$
    \STATE Update $\mathbf{m}_h^{(i+1)} = \gamma_1 \mathbf{m}_h^{(i)} + (1-\gamma_1) \frac{\partial f}{\partial \mathbf{C}_h^{(i)}}$
    \STATE Update $\mathbf{m}_j^{(i+1)} = \gamma_1 \mathbf{m}_j^{(i)} + (1-\gamma_1) \frac{\partial f}{\partial \mathbf{C}_j^{(i)}}$
    \STATE Update $\mathbf{v}_h^{(i+1)} = \gamma_2 \mathbf{v}_h^{(i)} + (1-\gamma_2) \frac{\partial f}{\partial \mathbf{C}_h^{(i)}} \circ \frac{\partial f}{\partial \mathbf{C}_h^{(i)}}$
    \STATE Update $\mathbf{v}_j^{(i+1)} = \gamma_2 \mathbf{v}_j^{(i)} + (1-\gamma_2) \frac{\partial f}{\partial \mathbf{C}_j^{(i)}} \circ \frac{\partial f}{\partial \mathbf{C}_j^{(i)}}$
    \STATE Bias correction $\mathbf{\hat m}_h^{(i+1)} = \frac{\mathbf{m}_h^{(i+1)}}{1-\gamma_1^{i+1}}$
    \STATE Bias correction $\mathbf{\hat m}_j^{(i+1)} = \frac{\mathbf{m}_j^{(i+1)}}{1-\gamma_1^{i+1}}$
    \STATE Bias correction $\mathbf{\hat v}_h^{(i+1)} = \frac{\mathbf{v}_h^{(i+1)}}{1-\gamma_2^{i+1}}$
    \STATE Bias correction $\mathbf{\hat v}_j^{(i+1)} = \frac{\mathbf{v}_j^{(i+1)}}{1-\gamma_2^{i+1}}$
    \STATE Update parameters: $\mathbf{C}_h^{(i+1)} = \mathbf{C}_h^{(i)} - \alpha \frac{\mathbf{\hat m}_h^{(i+1)}}{\sqrt{\mathbf{\hat v}_h^{(i+1)}} + \mathbf{e}_h}$
    \STATE Update parameters: $\mathbf{C}_j^{(i+1)} = \mathbf{C}_j^{(i)} - \alpha \frac{\mathbf{\hat m}_j^{(i+1)}}{\sqrt{\mathbf{\hat v}_j^{(i+1)}} + \mathbf{e}_j}$
    \STATE Update $i = i + 1$
\ENDWHILE
\RETURN each pair of parameters $\mathbf{C}_h$ and $\mathbf{C}_j$
\end{algorithmic}
\end{algorithm}
The estimators $\mathbf{\hat C}_1,...,\mathbf{\hat C}_n$ are obtained by Algorithm 1 and consequently the optimal $p$-dimensional FMDS configurations are derived by $\mathbf{\hat x}_1(t) = \mathbf{\hat C}_1 \boldsymbol{\beta}(t),...,\mathbf{\hat x}_n(t) = \mathbf{\hat C}_n \boldsymbol{\beta}(t)$.
\section{Simulation Study}
\noindent To evaluate the performance of the proposed FMDS model, we conduct a simulation study by assessing the goodness of the approximation of the estimators as the sample size of time period increases. As one of the measurement method to reveal how the FMDS model fits data well, the root-mean-square error ($RMSE$) is used to compare estimated dissimilarities and observed dissimilarities in the simulations and expect to see a smaller value of $RMSE$ when the sample size of time period increases in different scenarios.

\subsection{Accuracy of Observed Dissimilarities and Estimated Dissimilarities}
Our goal is to construct a $p$-dimensional FMDS representation $\mathbf{x}_1(t),...,\mathbf{x}_n(t) \in \mathbb{R}^p$ for $n$ functional objects $\mathbf{y}_1(t),...,\mathbf{y}_n(t) \in \mathbb{R}^r$, where $p<r$. We take $n=50$ and $p=2$ as default in the simulation setting to compare the RMSE of the observed dissimilarities $\|\mathbf{y}_i(t) - \mathbf{y}_j(t)\|$ and the estimated dissimilarities $\|\mathbf{\hat x}_i(t) - \mathbf{\hat x}_j(t)\|$, where $i,j=1,...,n$ and $i \neq j$. To obtain the synthetic functional data, we let each $\mathbf{y}_i(t) = \mathbf{C}_i^{(y)} \boldsymbol{\beta}(t)$, for $i=1,...,n$, where $\boldsymbol{\beta}(t)$ is a vector of $q$ B-spline basis functions and $\mathbf{C}_i^{(y)}$'s are $(p \times q)$ coefficient matrices.

We can start generating the coefficient matrices $\mathbf{C}_i^{(y)}$'s from a Gaussian. Let $\mathbf{v}_i^{(y)}$ be $vec(\mathbf{C}_i^{(y)})$ and be independent and identically distributed by a multivariate normal distribution $MN(\mathbf{0, \mathbf{\Sigma}})$, where $\mathbf{\Sigma}$ is a $(pq \times pq)$ identity matrix so that we can form the coefficient matrices $\mathbf{C}_1^{(y)},...,\mathbf{C}_n^{(y)}$. As for the basis function $\boldsymbol{\beta}(t)$, we choose the cubic B-spline families with $L=5$ and $L=10$ equally spaced knots, respectively. Accordingly, $q=5+4=9$ and $q=10+4=14$ for two individual models. In each model, we consider four sample sizes for the time period $m$: 15, 50, 100, 200. Thus, the simulation performs a total of 8 scenarios. Each scenario is replicated 300 times, which means that we run 300 times of the random sample of $\mathbf{C}_1^{(y)},...,\mathbf{C}_n^{(y)}$.

Given that the dissimilarities are measured by the Euclidean distances, the target function is derived as 
\begin{align*}
F(\mathbf{C}_1,...,\mathbf{C}_n) &= \sum_{i=1}^{n-1} \sum_{j=i+1}^{n} \sum_{k=1}^{m} [\|\mathbf{y}_i(t_k) - \mathbf{y}_j(t_k)\|^2 - \|\mathbf{x}_i(t_k) - \mathbf{x}_j(t_k)\|^2]^2 \\
 &= \sum_{i=1}^{n-1} \sum_{j=i+1}^{n} \sum_{k=1}^{m} [\|(\mathbf{C}_i^{(y)} \boldsymbol{\beta}(t_k) - \mathbf{C}_j^{(y)} \boldsymbol{\beta}(t_k)\|^2 - \|(\mathbf{C}_i \boldsymbol{\beta}(t_k) - \mathbf{C}_j \boldsymbol{\beta}(t_k)\|^2]^2,
\end{align*}
We do not use a roughness penalty in the above target function because there are only five or ten knots on the B-splines. When the number of knots is not very large, the variance of the estimators is not large. Thus, we do not need a roughness penalty to avoid the overfitting issue. 

Deploying the Algorithm 1: Modified Adam Stochastic Gradient Descent (SGD) for FMDS introduced in Section 2, we have $\mathbf{\hat C}_i$'s estimated by locally minimizing the above target function $F(\mathbf{C}_1,...,\mathbf{C}_n)$. Denote $d_{ij}(t) = \|\mathbf{y}_i(t) - \mathbf{y}_j(t)\|$ as the observed dissimilarity and let $\hat{d}_{ij}(t) = \|\mathbf{\hat x}_i(t) - \mathbf{\hat x}_j(t)\| = \|\mathbf{\hat{C}}_i \boldsymbol{\beta}(t) - \mathbf{\hat{C}}_j \boldsymbol{\beta}(t)\|$ be the estimated FMDS dissimilarity between the objects $i$ and $j$. For the time periods $m=15, 50, 100, 200$, the mean squared error ($MSE$) of $d_{ij}(t_k)$ and $\hat{d}_{ij}(t_k)$ for each scenario in one replication can be determined as 
\begin{align*}
MSE(m)_r = \frac{\sum_{i=1}^{n-1} \sum_{j=i+1}^{n} \sum_{k=1}^{m} (d_{ij}(t_k) - \hat{d}_{ij}(t_k))^2}{mn(n-1)/2}, \tag{3.1}
\end{align*}
for $r=1,2,...,300$. After we run 300 replications, the $RMSE$ of $d_{ij}(t_k)$ and $\hat{d}_{ij}(t_k)$ for each $m$ is evaluated as 
\begin{align*}
RMSE(m) = \sqrt{\frac{\sum_{r=1}^{300} MSE(m)_r}{300}}. \tag{3.2}
\end{align*}
We expect to see that $RMSE(m)$ will decrease as the value of $m$ increases.

\subsection{Goodness of Estimators}
On the other hand, we want to assess the goodness of the parameters $\mathbf{C}_i's$, for $i=1,...,n$. Since the estimated FMDS dissimilarities $\hat{d}_{ij}(t)$'s are rotation and translation invariant, the optimal solutions $\mathbf{\hat x}_i(t) = \mathbf{\hat C}_i \boldsymbol{\beta}(t)$ are rotation and translation equivalent. For any $p \times p$ orthogonal matrix $\boldsymbol{\Gamma}$, we have 
\begin{align*}
\hat{d}_{ij}(t) = \|\mathbf{\hat x}_i(t) - \mathbf{\hat x}_j(t)\| = \|\boldsymbol{\Gamma} \mathbf{\hat x}_i(t) - \boldsymbol{\Gamma} \mathbf{\hat x}_j(t)\|,
\end{align*}
hence, the $\{\boldsymbol{\Gamma} \mathbf{\hat x}_i(t)\}$'s provide an approximation to the dissimilarities $d_{ij}(t)$'s as good as the $\mathbf{\hat x}_i(t)$'s produce, that is, all $\{\boldsymbol{\Gamma} \mathbf{\hat C}_i \boldsymbol{\beta}(t)\}$'s are equivalent as the optimal FMDS solutions $\mathbf{\hat x}_i(t) = \mathbf{\hat C}_i \boldsymbol{\beta}(t)$, for $i=1,...,n$. Accordingly, our another purpose is to find the optimal orthogonal matrix $\boldsymbol{\hat{\Gamma}}$ in each replication such that 
\begin{align*}
\mathcal{G}(\boldsymbol \Gamma) &= \sum_{i=1}^n \int_{t=1}^m \|\boldsymbol{\Gamma} \mathbf{\hat C}_i \boldsymbol{\beta}(t) - \mathbf{C}_i^{(y)} \boldsymbol{\beta}(t)\|^2 \ dt\\
 &= \sum_{i=1}^n \int_{t=1}^m (\boldsymbol{\Gamma} \mathbf{\hat C}_i \boldsymbol{\beta}(t) - \mathbf{C}_i^{(y)} \boldsymbol{\beta}(t))^{\top} (\boldsymbol{\Gamma} \mathbf{\hat C}_i \boldsymbol{\beta}(t) - \mathbf{C}_i^{(y)} \boldsymbol{\beta}(t))\ dt,
\end{align*}
arrives at the minimum. It can be done by implementing a curvilinear search method with the Barzilai-Borwein (BB) step size which was proposed by \cite{wen2013feasible}. In order to facilitate this algorithm to find the optimization with the orthogonality constraint $\boldsymbol{\Gamma}^{\top} \boldsymbol{\Gamma} = \mathbf{I}$ on RStudio conveniently, we apply the trapezoidal rule to $\mathcal{G}$ so that it becomes 
\begin{align*}
\mathcal{G} (\boldsymbol \Gamma) &= \sum_{i=1}^n \{\frac{0.5}{2}[g(1) + 2g(2) + 2g(3) + ... +2g(2m-2) + g(2m-1)]\} \\
&= \frac{1}{4} \sum_{i=1}^n [g(1) + 2g(2) + 2g(3) + ... +2g(2m-2) + g(2m-1)],
\end{align*}
where $g(k) = (\boldsymbol{\Gamma} \mathbf{\hat C}_i \boldsymbol{\beta}(t_k) - \mathbf{C}_i^{(y)} \boldsymbol{\beta}(t_k))^{\top} (\boldsymbol{\Gamma} \mathbf{\hat C}_i \boldsymbol{\beta}(t_k) - \mathbf{C}_i^{(y)} \boldsymbol{\beta}(t_k))$. The gradient function with respect to $\boldsymbol{\Gamma}$ is required in this method, and it is derived as 
\begin{align*}
\frac {\partial \mathcal{G}}{\partial \boldsymbol \Gamma} &=
\frac {1}{4} \sum_{i=1}^n (2 \mathbf I)(\boldsymbol{\Gamma} \mathbf{\hat C}_i \boldsymbol{\beta}(t_1) - \mathbf{C}_i^{(y)} \boldsymbol{\beta}(t_1))(\mathbf{\hat C}_i \boldsymbol{\beta}(t_1))^{\top} \\
&+ \frac{1}{2} \sum_{i=1}^n \sum_{k=2}^{2m-2}(2 \mathbf I)(\boldsymbol{\Gamma} \mathbf{\hat C}_i \boldsymbol{\beta}(t_k) - \mathbf{C}_i^{(y)} \boldsymbol{\beta}(t_k))(\mathbf{\hat C}_i \boldsymbol{\beta}(t_k))^{\top} \\
&+ \frac {1}{4} \sum_{i=1}^n (2 \mathbf I)(\boldsymbol{\Gamma} \mathbf{\hat C}_i \boldsymbol{\beta}(t_{2m-1}) - \mathbf{C}_i^{(y)} \boldsymbol{\beta}(t_{2m-1}))(\mathbf{\hat C}_i \boldsymbol{\beta}(t_{2m-1}))^{\top}
\end{align*}
Since the algorithm is an iterative method, the initial guess for the $p \times p$ orthogonal matrix $\boldsymbol{\Gamma}_0$ is needed and generated by the standard Gaussian. If $\boldsymbol{\Gamma}_0$ is not orthogonal, it will be processed by Gram-Schmidt. According to Lemma 1 in \cite{wen2013feasible}, let 
\begin{align*}
\mathbf G = \frac{\partial \mathcal {G}}{\partial \boldsymbol \Gamma}.
\end{align*}

We also define 
\begin{center}
$\nabla \mathcal{G} = \mathbf G - \boldsymbol{\Gamma} \mathbf{G}^{\top} \boldsymbol{\Gamma}$ and $\mathbf A = \mathbf G \boldsymbol \Gamma^{\top} - \boldsymbol \Gamma \mathbf G^{\top}$
\end{center}
which leads to $\nabla \mathcal{G} = \mathbf{A} \boldsymbol{\Gamma}$. Meanwhile, there are several parameters needed in the method, including $\rho_1, \delta, \eta, \epsilon \in (0,\ 1)$. In the $k^{th}$ iteration, we set $\tau_k$ to either
\begin{align*}
\tau_k = \frac{tr((S_{k-1})^{\top} S_{k-1})}{|tr((S_{k-1})^{\top} Y_{k-1})|} \quad \text{or} \quad \tau_k = \frac{|tr((S_{k-1})^{\top} Y_{k-1})|}{tr((Y_{k-1})^{\top} Y_{k-1})},
\end{align*}
where $S_{k-1} = \boldsymbol \Gamma_k - \boldsymbol \Gamma_{k-1}$ and $\mathbf Y_{k-1} = \nabla \mathcal{G}(\boldsymbol \Gamma_k) - \nabla \mathcal{G}(\boldsymbol \Gamma_{k-1})$. The new orthogonal matrix $\boldsymbol{\Gamma}_{k+1}$ is generated iteratively by the steps in a Curvilinear Search method with BB steps introduced by \cite{wen2013feasible}. The iteration will stop when $\|\nabla \mathcal{G}(\boldsymbol \Gamma_k)\| \leq \epsilon$ and we will have the optimal orthogonal matrix $\boldsymbol{\hat{\Gamma}}$ which minimizes $\mathcal{G} (\boldsymbol \Gamma)$. 

Based on the optimal equivalent solutions $\{\boldsymbol{\hat {\Gamma}} \mathbf{\hat C}_i \boldsymbol{\beta}(t)\}$'s in each replication, for $i=1,...,n$, we are also able to see how the parameter estimators $\mathbf{C}_1,...,\mathbf{C}_n$ perform in the model by computing the $RMSE$ of $\boldsymbol{\hat{\Gamma}} \mathbf{\hat C}_i$ and $\mathbf{C}_i^{(y)}$ through the 300 replications with five and ten knots, respectively.  It can be done as follows. First of all, for each scenario in the replication $r$, where $r=1,...,300$, we evaluate the $MSE$ of $\boldsymbol{\hat{\Gamma}} \mathbf{\hat C}_i$ and $\mathbf{C}_i^{(y)}$ with 
\begin{align*}
MSE(\boldsymbol{\hat{\Gamma}}, m)_r = \frac{\sum_{i=1}^n \sum_{j=1}^{pq} (\hat{a}_j^{(i)} - a_j^{(i)})^2}{npq}, \tag{3.3}
\end{align*}
where $\hat{a}_j^{(i)}$'s are the elements of $vec(\boldsymbol{\hat{\Gamma}} \mathbf{\hat C}_i)$, and $a_j^{(i)}$'s are the elements of $vec(\mathbf{C}_i^{(y)})$. After we run 300 replications, we will have the $RMSE$ of $\boldsymbol{\hat{\Gamma}} \mathbf{\hat C}_i$ and $\mathbf{C}_i^{(y)}$ for each $m$ being evaluated as 
\begin{align*}
RMSE(\boldsymbol{\hat{\Gamma}}, m) = \sqrt{\frac{\sum_{r=1}^{300} MSE(\boldsymbol{\hat{\Gamma}}, m)_r}{300}}. \tag{3.4}
\end{align*}
Again, we expect to see that $RMSE(\boldsymbol{\hat{\Gamma}},m)$ will decrease as the value of $m$ increases.

\subsection{Simulation Results}
Based on the calculation of $RMSE(m)$ in (3.1) and (3.2), we have Table A which shows that the $RMSE$ of $d_{ij}(t_k)$ and $\|\mathbf{\hat x}_i(t) - \mathbf{\hat x}_j(t)\|$ decreases as $m$ increases, as expected, when the number of knots $L = 5$ on the cubic B-splines $\boldsymbol{\beta}(t)$. Meanwhile, computing $RMSE(\boldsymbol{\hat{\Gamma}},m)$ in (3.3) and (3.4) for the performance of the estimators $\mathbf{C}_i$'s, we can see that the $RMSE$ of $\mathbf{C}_i^{(y)}$ and $\boldsymbol{\hat{\Gamma}} \mathbf{\hat{C}}_i$ also decreases as $m$ increases. Comparing the values of $RMSE$ with $m=100$ and $m=200$, we have the $RMSE$ values for both estimators $\|\mathbf{x}_i(t) - \mathbf{x}_j(t)\|$ and $\mathbf{C}_i$ decrease faster from the change of $m=15$ to $m=50$. The values of $RMSE$ tend to relatively stable when $m>50$, but also continue to decrease as $m$ increases, as expected.

\begin{table}[htbp]
  \captionsetup{labelformat=empty}
  \centering
  \begin{tabular}{|c|c|c|c|c|}
    \hline
    \textbf{Estimator} & $m=15$ & $m=50$ & $m=100$ & $m=200$ \\
    \hline
    $\|\mathbf{x}_i(t) - \mathbf{x}_j(t)\|$ & 2.198 & 1.254 & 1.050 & 0.972 \\
    \hline
    $\mathbf{C}_i$ & 0.399 & 0.306 & 0.279 & 0.266 \\
    \hline
  \end{tabular}
  \caption{\textbf{Table A:} $RMSE(m)$ and $RMSE(\boldsymbol{\hat{\Gamma}},m)$ with $m=15, 50, 100, 200$ and $L=5$}
\end{table}

We should expect the same trend of the $RMSE$ values in the situation with $L=10$ knots on the cubic B-splines $\boldsymbol{\beta}(t)$, shown in Table B. The values of $RMSE$ drop rapidly when $m=15$ changes to $m=50$, then tend to relatively stable when $m>50$. The situation with $L=10$ knots on the cubic B-splines $\boldsymbol{\beta}(t)$ makes the estimators even more accurate, especially for the estimators $\mathbf{C}_i$'s, resulting in the smaller values of $RMSE$ compared to the values in Table A. This is attributed to the smaller bias as the number of knots increases.

\begin{table}[htbp]
  \captionsetup{labelformat=empty}
  \centering
  \begin{tabular}{|c|c|c|c|c|}
    \hline
    \textbf{Estimator} & $m=15$ & $m=50$ & $m=100$ & $m=200$ \\
    \hline
    $\|\mathbf{x}_i(t) - \mathbf{x}_j(t)\|$ & 2.245 & 0.925 & 0.628 & 0.514 \\
    \hline
    $\mathbf{C}_i$ & 0.622 & 0.257 & 0.215 & 0.211 \\
    \hline
  \end{tabular}
  \caption{\textbf{Table B:} $RMSE(m)$ and $RMSE(\boldsymbol{\hat{\Gamma}},m)$ with $m=15, 50, 100, 200$ and $L=10$}
\end{table}

Overall, the decreasing values of $RMSE(m)$ and $RMSE(\boldsymbol{\hat{\Gamma}},m)$ in Table A and Table B validate the effectiveness of the FMDS model in time-varying dissimilarities. Our experiments confirm that the model delivers robust performance on simulated data, especially in the situation with $L=10$ knots on the cubic B-splines $\boldsymbol{\beta}(t)$. In addition, by utilizing the modified Adam stochastic gradient descent method, we perform the nonlinear optimization 51 times faster, verifying the efficiency of the modified Adam SGD algorithm for the large-scale optimization in the FMDS model.
\section{Real-Data Application}
\noindent Motivated by the example of the variant classical MDS in our previous work (\cite{li2025functional}) for the daily closing prices of the stocks in a given week, we consider the daily closing prices of the 500 stocks that make up the S\&P 500 Index in the year 2018 and apply the FMDS model to the time-varying dissimilarities of the daily closing prices. The data is publicly available online at the Yahoo Finance website, https://finance.yahoo.com.

Suppose that there are $r$ trading days in the month $t$ and denote the daily closing prices for each stock in the month as $\mathbf{y}_i(t) = [y_{i1}(t),...,y_{ir}(t)]^\top$, for $i=1,...,500$. To be clarified, $\mathbf{y}_i(t)$ is not a function of $t$, but it only stands for the daily closing prices of the $i^{th}$ stock during the month $t$. Accordingly, the correlation of stocks $i$ and $j$ in the months $t$ is given by 
\begin{align*}
R_{ij}(t) = \frac{\sum_{k=1}^{r}(y_{ik}(t)-\bar{y}_i(t))(y_{jk}(t)-\bar{y}_j(t))}{\sqrt{\sum_{k=1}^{r}(y_{ik}(t)-\bar{y}_i(t))^2} \sqrt{\sum_{k=1}^{r}(y_{jk}(t)-\bar{y}_j(t))^2}} ,
\end{align*}
for $i,j=1,...,500$, where $\bar{y}_i(t) = r^{-1} \sum_{k=1}^{r} y_{ik}(t)$ and $\bar{y}_j(t) = r^{-1} \sum_{k=1}^{r} y_{jk}(t)$. Instead of using the Euclidean distances, we define the dissimilarities of the closing prices as
\begin{align*}
d_{ij}(t) = \frac{1-R_{ij}(t)}{2}.
\end{align*}
Since $-1 \leq R_{ij}(t) \leq 1$, each $d_{ij}(t)$ satisfies $0 \leq d_{ij}(t) \leq 1$. If we do this for 12 months $t_1,...,t_{12}$, we will obtain a series of dissimilarities matrices
\begin{align*}
\mathbf{D}(t_k) = 
\begin{bmatrix}
0 & d_{12}(t_k) & \cdots & d_{1,500}(t_k) \\
d_{21}(t_k) & 0 & \cdots & d_{2,500}(t_k) \\
\vdots & \vdots & \ddots & \vdots \\
d_{500,1}(t_k) & d_{500,2}(t_k) & \cdots & 0
\end{bmatrix},
\end{align*}
where
$ 
\begin{cases}
d_{ij}(t_k) = d_{ji}(t_k), & \text{for } i \neq j, \text{ } \text{and } k=1,...,12,\\
d_{ii}(t_k) = 0, & \text{for all } i, \text{ } \text{and } k=1,...,12.
\end{cases}
$

Since each $\mathbf{D}(t_k)$ is a symmetric matrix with zeroes in the diagonal, we can vectorize the upper triangular part of $\mathbf{D}(t_k)$ for each $k$ without the diagonal for convenience so that we form a super dissimilarity matrix $\mathbf{D}$ with these vectorized upper triangular parts for the 12 months. As a result, it is constituted by $C_{500}^2 = 124750$ rows for the combinations of two stocks and 12 columns for the months, as follows.
\begin{align*}
\mathbf{D} = 
\begin{bmatrix}
d_{12}(t_1) & d_{12}(t_2) & \cdots & d_{12}(t_{12}) \\
d_{13}(t_1) & d_{13}(t_2) & \cdots & d_{13}(t_{12}) \\
d_{23}(t_1) & d_{23}(t_2) & \cdots & d_{23}(t_{12}) \\
d_{14}(t_1) & d_{14}(t_2) & \cdots & d_{14}(t_{12}) \\
d_{24}(t_1) & d_{24}(t_2) & \cdots & d_{24}(t_{12}) \\
d_{34}(t_1) & d_{34}(t_2) & \cdots & d_{34}(t_{12}) \\
\vdots & \vdots & \ddots & \vdots \\
d_{499,500}(t_1) & d_{499,500}(t_2) & \cdots & d_{499,500}(t_{12})
\end{bmatrix}.
\end{align*}

Now our goal is to come up with a method to obtain a smoothly time-varying MDS representation $\mathbf{x}_1(t),...,\mathbf{x}_{500}(t) \in \mathbb{R}^p$ $(p < r)$ for the stocks of the S\&P 500 Index with each $t$, such that
\begin{align*}
\|\mathbf{x}_i(t) - \mathbf{x}_j(t)\| \approx d_{ij}(t),
\end{align*}
for $i=1,...,499$, $j=1,...500$ and $i<j$. Assume that $\mathbf{x}_i(t) = \mathbf{C}_i \boldsymbol{\beta}(t)$, where $\boldsymbol{\beta}(t)$ is a vector of $q$ cubic B-spline basis functions, and each $\mathbf{C}_i$ is a $(p \times q)$ coefficient matrix. At the end of the case study, we display the FMDS representation $\mathbf{x}_1(t),...,\mathbf{x}_{500}(t)$ in a 2-dimensional map in order to have a convenient visualization. That is, $p=2$. Also, we select the first 15 stocks to analyze their clusters in January, February, March and April.

Considering that $\mathbf{D}=[d_{ij}(t_k)]$ is a $(124750 \times 12)$ super dissimilarity matrix and each $\mathbf{x}_i(t)$ is 2-dimensional, we choose a smaller value of $q$ so that we can avoid the time-consuming optimization. On the other hand, it is also unacceptable if $q=m=12$, because it leads to zero bias of the estimators but very high variance of the estimators. Hence, as a comprise between the bias and variance trade-off, we use $q=10$, i.e. $L=6$ on the cubic B-splines. Consequently, each $\mathbf{C}_i$ is a $(2 \times 10)$ coefficient matrix and $\boldsymbol{\beta}(t)$ is a vector of 10 cubic B-spline basis functions with 6 equally spaced knots in the interval $[1, 12]$.

Since there are only 6 knots used in the basis functions, it is not necessary to add a roughness penalty to the least-square criterion. According to the target function (2.1) in Section 2, we can estimate the $\mathbf{C}_i$'s by minimizing
\begin{align*}
F(\mathbf{C}_1,...,\mathbf{C}_{500}) &= \sum_{i<j}^{500} \sum_{k=1}^{12} [d_{ij}^2(t_k) - \|\mathbf{x}_i(t_k) - \mathbf{x}_j(t_k)\|^2]^2 \\
 &= \sum_{i<j}^{500} \sum_{k=1}^{12} [d_{ij}^2(t_k) - \|\mathbf{C}_i \boldsymbol{\beta}(t_k) - \mathbf{C}_j \boldsymbol{\beta}(t_k)\|^2]^2 \\
 &= \sum_{i=1}^{499} \sum_{j=i+1}^{500} \sum_{k=1}^{12} [d_{ij}^2(t_k) - (\mathbf{C}_i \boldsymbol{\beta}(t_k) - \mathbf{C}_j \boldsymbol{\beta}(t_k))^\top (\mathbf{C}_i \boldsymbol{\beta}(t_k) - \mathbf{C}_j \boldsymbol{\beta}(t_k))]^2. \tag{4.1}
\end{align*}

Subsequently, we use the modified Adam SGD method in the Algorithm 1 to solve the above optimization problem. As a result, the closing price of each stock can be modeled as a smooth function of time $t$, which is expressed as
\begin{align*}
\mathbf{\hat x}_i(t) = \mathbf{\hat C}_i \boldsymbol{\beta}(t),
\end{align*}
for $i=1,...,500$. Figure 1 shows the map of 2-dimensional FMDS solutions $\mathbf{\hat x}_1(t),...,\mathbf{\hat x}_{500}(t)$ for the stocks in the 12 months of the year 2018. Each graph is for one month. We can see that some stocks are clustered in January, February, March, October and December but many stocks are dispersed in the other months. \\

\begin{figure}[H]
    \centering
    \includegraphics[width=1.0\linewidth]{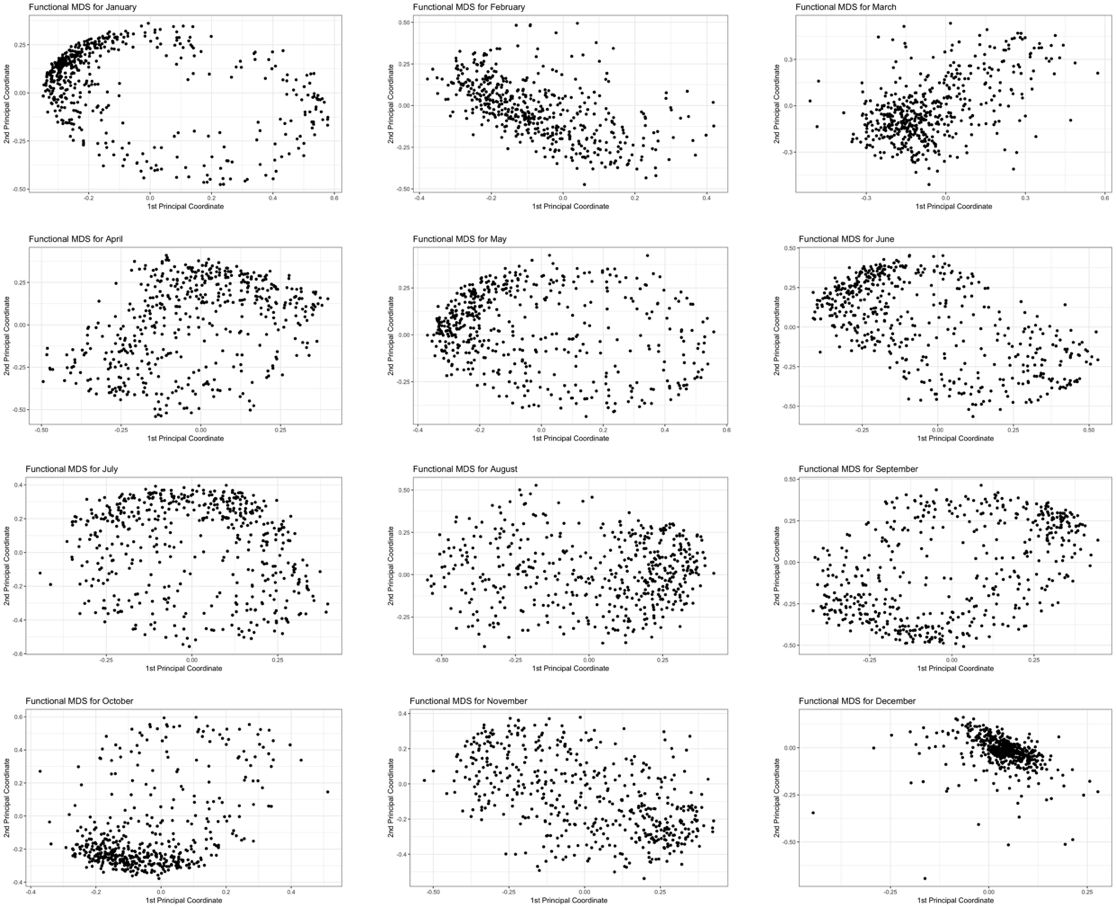}
    \captionsetup{font=small}
    \caption{2D FMDS Maps of the S\&P Stocks Closing Prices in the 12 Months of Year 2018}
    \label{fig:enter-label}
\end{figure}

Specifically, we select the first 15 stocks for 4 months as an example to see how the stocks move. Figure 2 shows that the 2D FMDS maps for 15 stocks: A, AAL, AAP, AAPL, ABBV, ABC, ABMD, ABT, ACN, ADBE, ADI, ADM, ADP, ADS, and ADSK in January (top left), February (top right), March (bottom left), and April (bottom right), respectively. \\

\begin{figure}[H]
    \centering
    \includegraphics[width=1.0\linewidth]{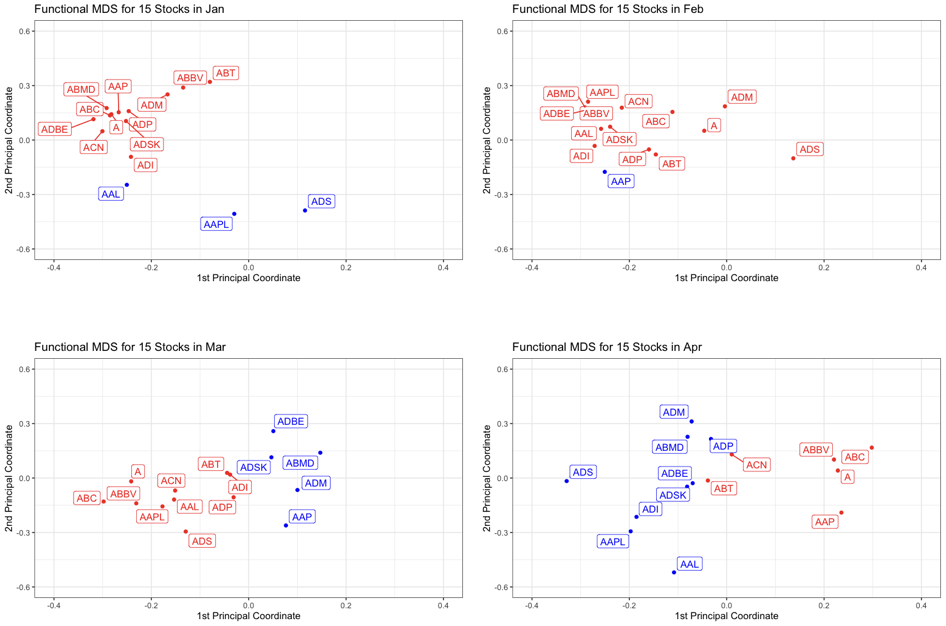}
    \captionsetup{font=small}
    \caption{FMDS for 15 Stocks Closing Prices in January, February, March, April}
    \label{fig:enter-label}
\end{figure}

Similar to some methods in cluster analysis, the stocks can be grouped in clusters based on some conditions. For example, we randomly select a stock as the center point to separate the stocks in two clusters. Let’s select the stock A (Agilent Technologies) as the center point. Denote $\hat{d}_{1j}(t)$ as the estimated FMDS dissimilarities between the stock A and the $j^{th}$ stock. The stocks in the red cluster have the dissimilarity (calculated as the Euclidean distance) $\hat{d}_{1j}(t) < 0.3$ from the stock A, whereas the stocks in the blue cluster have the dissimilarity $\hat{d}_{1j}(t) \geq 0.3$ from the stock A. In other words, the stock A is closer to the stocks in the red cluster than the stocks in the blue cluster. Meanwhile, we can see how the stocks move in each month, for instance, the stocks A and AAP (Advance Auto Parts) are close to each other in January, but they move apart from each other in February and March. In contrast, the stocks A and AAPL (Apple) belong to different clusters in January, but they move closer in February and March. On the other hand, the stock A, ABBV (AbbVie) and ABC (AmerisourceBergen) are very close to each other. The small dissimilarities among these three stocks demonstrate that they have an impact on each other during these four months. It makes sense because all of them belong to the sector in Health Care. Consequently, the maps made by the FMDS model show that the clusters have different stocks in each month.

Regarding the smoothness, Figure 3 shows two examples for the comparison between the observed dissimilarity $d_{ij}(t)$ (blue and dashed lines) and the estimated FMDS dissimilarity $\hat{d}_{ij}(t)$ (red and smooth curves). The left panel reflects the example of the pair stocks of A and AAP, and the right panel shows the example of the pair stocks of AAP and BA (Boeing). According to the red curves in the panels, we can see how the dissimilarity of A and AAP and the dissimilarity of AAP and BA change in month smoothly. Moreover, the red smooth curves are very close to the corresponding blue dashed lines. It means that the 2D smoothly time-varying representations $\mathbf{\hat{x}}_i(t) = \mathbf{\hat C}_i \boldsymbol{\beta}(t)$ are well modeled in FMDS without changing the trend of the observed dissimilarities, preserving the original dissimilarities. \\

\begin{figure}[H]
    \centering
    \includegraphics[width=1.0\linewidth]{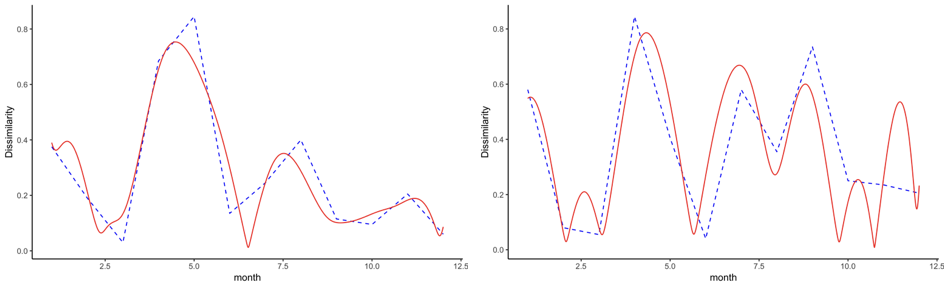}
    \captionsetup{font=small}
    \caption{Dissimilarities Change Smoothly in Months}
    \label{fig:enter-label}
\end{figure}

Alternatively, we create the Shepard diagrams in Figure 4 that assess how well the estimated FMDS dissimilarities preserve the observed dissimilarities between the daily closing prices of the stocks in each month. The graphs suggest that most of the estimated FMDS dissimilarity $\|\mathbf{\hat x}_i(t) - \mathbf{\hat x}_j(t)\|$ tend to close to the observed dissimilarity $d_{ij}(t)$, especially in January, February, March, October, November and December. This finding is valuable for a deep dive in the stock market throughout the year 2018. The FMDS model exhibits promising performance in these 6 months (the beginning and end of the year) compared to the other months in the middle of the year. Several plausible factors may lead to this pattern. For example, it is common that there are stronger market-wide movements at the beginning and end of the year with more uniform responses across the stocks. It may induce higher inter-stock correlations, leading to a more stable dissimilarity structure. However, the stock market experienced an unstable situation driven by some major political and economic decisions during the mid-year periods of 2018, which probably affected the predictions in FMDS. We will continue this discussion in Section 5. \\

\begin{figure}[H]
    \centering
    \includegraphics[width=1.0\linewidth]{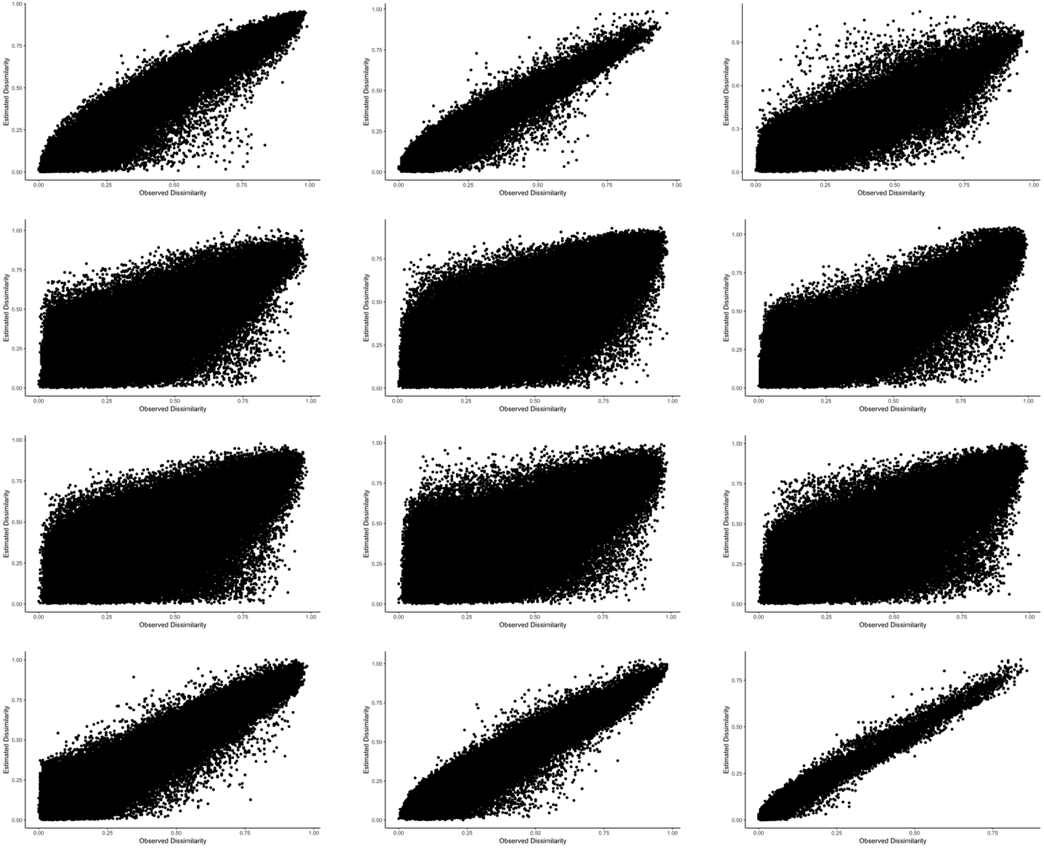}
    \captionsetup{font=small}
    \caption{The Shepard Diagrams for Each Month in Year 2018}
    \label{fig:enter-label}
\end{figure}

Last but not least, Figure 5 shows the residuals from the fit plotted against the sequence number, where residual is $\|\mathbf{\hat x}_i(t_k) - \mathbf{\hat x}_j(t_k)\| - d_{ij}(t_k)$. Although the residuals are relatively large for some stocks, the graph implies that the residuals are not large for most of the stocks. Hence, the difference between the observed and estimated FMDS dissimilarities tend to zero in general. It satisfies the situation in Figure 4. As a conclusion, there are 70.5\% of the stocks having the absolute residual smaller than or equal to 0.1. \\

\begin{figure}[H]
    \centering
    \includegraphics[width=1.0\linewidth]{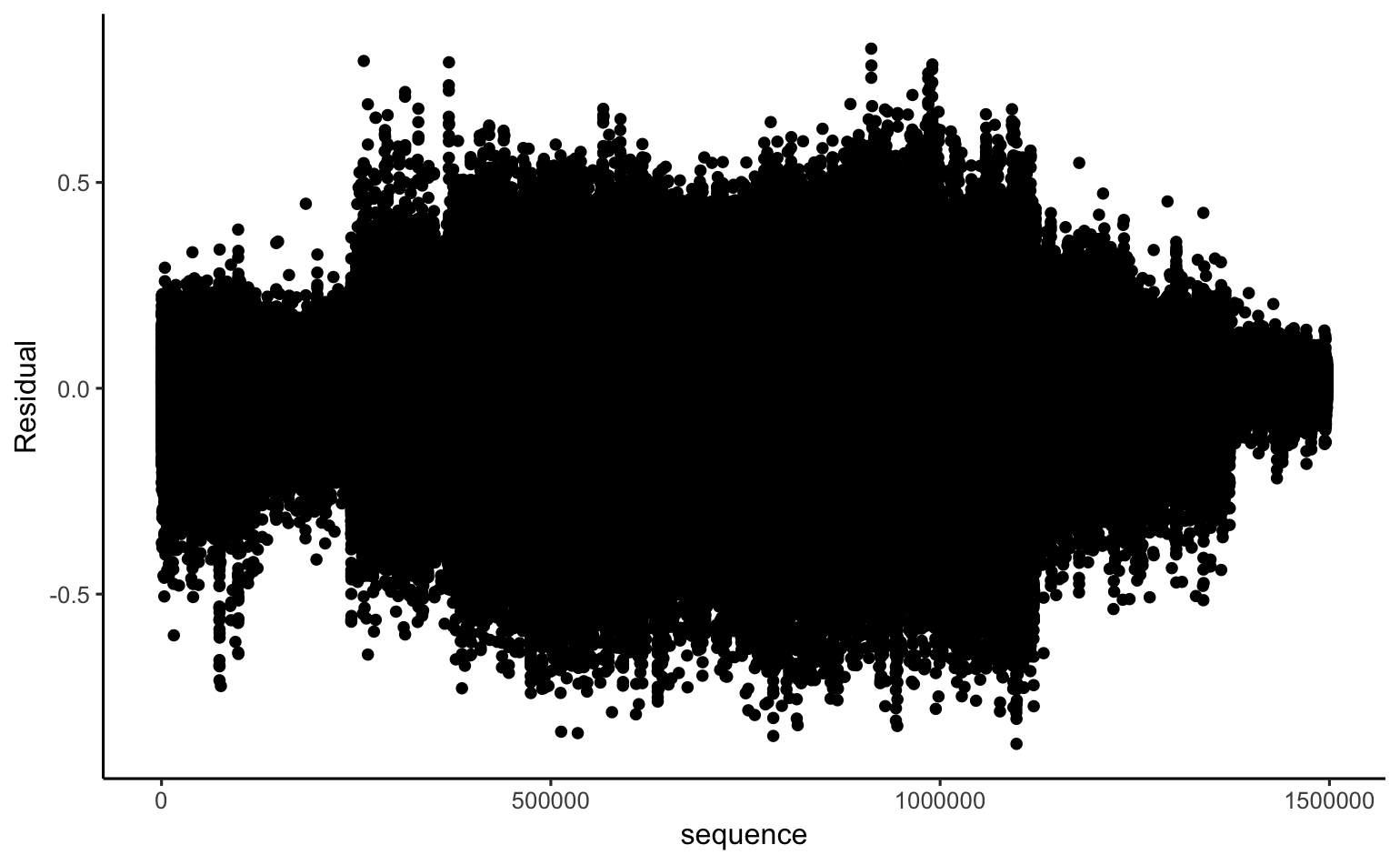}
    \captionsetup{font=small}
    \caption{The Residuals of $\|\mathbf{\hat x}_i(t_k) - \mathbf{\hat x}_j(t_k)\|$ and $d_{ij}(t_k)$}
    \label{fig:enter-label}
\end{figure}
\section{Discussion}
\noindent The simulation study and real-data application demonstrate that the FMDS model not only preserves dissimilarities of objects in dimensional reduction but also improves the smoothness in constructing a low-dimensional solution when dissimilarities change over time. Despite its effectiveness in practical applications, there are inevitable challenges in the optimization shown in Section 4. For example, the dissimilarities are well estimated at the beginning and end of the year, while they are underestimated or overestimated in the middle of the year.

In fact, this happens as we expect, due to four main reasons for it: (1) As we mentioned in Section 4, many factors in reality may cause underestimated or overestimated dissimilarities during the mid-year periods, including economic policy shifts and geopolitical events. (2) We are using the modified Adam SGD algorithm, which updates the pair of estimators $\mathbf{C}_i$ and $\mathbf{C}_j$ randomly in each iteration, so there are probably not enough iterations to update some pairs. (3) We seek for the local minimum of the target function (4.1) instead of the global minimum, which is the essence and advantage of Adam SGD. Although global optimization methods may identify more optimal solutions in theory, they are often infeasible in high-dimensional problems and computationally prohibitive in practice. In contrast, local optimization enables faster convergence and scalability in high-dimensional data, offering a more efficient solution. This trade-off is especially advantageous in practical applications such as financial analysis, climate modeling, medical health research, or other real cases where timely and efficient analysis is crucial. (4) The stopping threshold $\epsilon$ in the iterations of the modified Adam SGD method is one of the reasons for different local minimums. The lower the stopping threshold, the smaller the difference between the estimated and observed dissimilarities. In this case study, the modified Adam SGD algorithm stops searching the local minimum of the target function (4.1) $F(\mathbf{C}_1,...,\mathbf{C}_{500})$ when all pairs of $h$ and $j$ have $\|\mathbf{C}_h^{(i+1)} - \mathbf{C}_h^{(i)}\| < 0.00075$ and $\|\mathbf{C}_j^{(i+1)} - \mathbf{C}_j^{(i)}\| < 0.00075$ as we set $\epsilon = 0.00075$ in the modified Adam SGD method.

While the modified Adam SGD method reduces the computation time from days to 5 minutes in the example of stock prices, there are limitations in the local computer. To address the challenges of the second and fourth reasons related to the number of iterations and the stopping threshold $\epsilon$, in the future we can use cloud computing services such as AWS that expand computing resources with more iterations and a smaller stopping threshold $\epsilon$. Also, by using cloud computing services, we can also conduct the numerical simulation with more trials to evaluate the consistency. Lastly, the FMDS model is not limited to use in the stock market. Assuming that a fraudulent customer is known in a credit card company, we can apply the FMDS model to find the other customers who have a small dissimilarity from this fraud so that it can help us detect more frauds. To extend the power of the FMDS model, we will do more experiments in other domains.
\section{Conclusion}
\noindent In this article, we study the consistency of the estimated solutions in the FMDS model through a comprehensive simulation and prove that the model performs well under Gaussian assumptions. Meanwhile, the modified Adam SGD method applied to the optimization of the estimators $\mathbf{C}_i$'s in this study efficiently reduces the computation burden. In the real-data application, we empirically validate the utility of the FMDS model in the stock market by presenting the time-varying dissimilarities change smoothly over the year and preserving the original dissimilarities as well. The results from the real case demonstrate that the model performs accurate relationships among the stocks and reveals temporal patterns and interpretable structures of the stocks that aligned with the market behaviors. Future work will extend the FMDS model to other domains such as fraud detection. 

\bibliographystyle{apalike}

\end{document}